\documentclass{iau}
\usepackage{graphicx} 

\title[IAUS291.~~Supernovae and compact objects] 
{Numerical modeling of core-collapse supernovae and compact objects} 

\author[Kohsuke Sumiyoshi]  
{Kohsuke Sumiyoshi$^1$
}

\affiliation{$^1$Numazu College of Technology, 
Ooka 3600, Numazu, Shizuoka 410-8501, Japan \\ email: {\tt sumi@numazu-ct.ac.jp} 
}

\pubyear{2012}
\volume{291}  
\jname{\mbox{Neutron Stars and Pulsars: Challenges and Opportunities after 80 years}}
\editors{J. van Leeuwen, ed.} 
\begin{document}

\maketitle

\begin{abstract}
Massive stars (M$\geq$ 10M$_{\odot}$) end their lives 
with spectacular explosions due to gravitational collapse.  
The collapse turns the stars into compact objects 
such as neutron stars and black holes 
with the ejection of cosmic rays and heavy elements.  
Despite the importance of these astrophysical events, 
the mechanism of supernova explosions has been an unsolved issue 
in astrophysics.  
This is because clarification of the supernova dynamics requires 
the full knowledge of nuclear and neutrino physics at extreme conditions, and 
large-scale numerical simulations of neutrino radiation hydrodynamics 
in multi-dimensions.  
This article is a brief overview of the understanding 
(with difficulty) of 
the supernova mechanism through the recent advance of numerical modeling 
at supercomputing facilities.  
Numerical studies with the progress of nuclear physics are applied 
to follow the evolution of compact objects with neutrino emissions 
in order to reveal the birth of pulsars/black holes from the massive stars.  
\end{abstract}


\firstsection 
\section{Introduction}

Supernova explosions are fascinating phenomena in the Universe 
(\cite[Bethe 1990]{Bethe1990}).  
First of all, they have bright displays by gigantic explosions, 
some of which were recorded even in the old Chinese literature, 
as in the case of SN 1054 (the Crab nebula).  
They are named as {\it guest stars}, which is the same expression 
for the description of SN 1054 noted in the old Japanese literature.  
Secondly, they are associated with bursts of neutrinos, 
as in the case of SN 1987A (\cite[Hirata et al. 1987]{Hirata1987}).  
In this event, neutrinos from the supernova were detected 
at terrestrial detectors.  
They revealed the importance of neutrinos in the supernova mechanism.  
This detection also led to the Nobel Prize in Physics in 2002 
(\cite[Koshiba 1992, 2003]{Koshiba1992}\cite[]{Koshiba2003}).  
Most importantly, 
the supernova is the birthplace of compact objects: 
neutron stars and black holes.  
It is essential to understand the supernova mechanism to know 
the distribution of compact objects from massive stars, 
together with the information on mass and rotation.  

Numerical modeling of core-collapse supernovae is the heart 
of the problem 
in order to grasp the timeline from massive stars to pulsars.  
Massive stars of $\sim$20M$_{\odot}$ evolves for $\sim$10$^7$ years 
with the formation of Fe core at the end of life.  
Gravitational collapse of the Fe core leads to 
core bounce, launch of shock wave and a resulting explosion 
in $\sim$1 s.  
At center, a proto-neutron star is born and evolves thermally 
by emitting supernova neutrinos for $\sim$20 s 
(e.g. \cite[Pons et al. 1999]{Pons99}).  
It further cools down by radiation for $\sim$10$^2$-10$^5$ years 
as a cold neutron star (\cite[e.g. Page et al. 2004]{Page2004}).  
What one wants to know is the connection 
between the stellar evolution and the neutron star modeling.  
To bridge a gap between massive stars and pulsars, 
one needs to perform the supernova simulations, 
which are the most difficult part of the problem.  
This article describes 
ongoing challenges to clarify the supernova explosions 
with multi-physics at extreme conditions in 3 dimensions 
toward the goal to reveal the birthplace of neutron stars.  

\section{Challenges to understand the supernova mechanism}

Despite the longstanding efforts over four decades, 
it is still not a simple task to reproduce supernova 
explosions in numerical modeling 
(\cite[Janka 2012]{Janka2012}, \cite[Kotake et al. 2012a]{kotake2012a}).  
Among many issues of the supernova modeling, 
three major aspects of the challenges are: 
exotic conditions, counteracting $\sim$1\% effects and multi-dimensions.  
To tackle these difficult problems, 
it is necessary to perform large-scale simulations 
on the latest supercomputers.  

To explain the current challenges in supernova physics, 
let us look into the details of supernova mechanism.  
Starting from the Fe core of the massive star, 
the gravitational collapse begins 
due to electron captures and photo-dissociations. 
The former reactions produce neutrinos, which escape freely at the beginning.  
These neutrinos are trapped inside the core at high densities during the collapse.  
Further collapse leads to the core bounce at just above the nuclear matter density, 
where the nuclear repulsive force is dominant.  
The shock wave is launched at the core bounce due to the stiffness of dense matter.  
If the shock wave propagates outwardly and reach the surface of the Fe core, 
this is the supernova explosion, 
which leaves the proto-neutron star and emits the supernova neutrinos.  
The proto-neutron star settles down as a cold neutron star afterward.  
The most difficult part among these stages is the way 
from the core bounce to the explosion 
with the three major challenges in the following subsections.  

\subsection{Nuclear physics at extreme conditions}

One of the key ingredients in the supernova modeling is 
nuclear physics at extreme conditions.  
It is mandatory to provide 
the properties of hot-dense matter (the equation of state) and 
the rates of neutrino reactions 
in the supernova core for numerical simulations.  
The density inside the supernova core becomes very high, 
being dense more than that in nuclei.  
The matter becomes very neutron-rich as compared with 
the stable nuclei such as $^{56}$Fe.  
The temperature may rise beyond $\sim$10$^{11}$ K ($\sim$10 MeV) 
in the central core, 
which is different from the cold neutron star case.  
In addition, it is necessary to consider 
the neutrino reactions under the influence of the hot-dense matter.  
One needs energy- and angle-dependent neutrino reaction rates for emission, 
absorption, scattering and pair processes.  
Neutrino processes via all targets including nucleons, nuclei and leptons 
must be considered (\cite[e.g. Burrows et al. 2006]{Burrows06npa}).  

As for the equation of state (EOS), it is a challenging task to provide 
the data set of EOS for wide range of density, temperature and composition 
for the supernova simulations.  
Although there are many results of the EOS for cold neutron stars, 
the set of EOS for supernovae have been limited before 
(\cite[Lattimer \& Swesty 1991]{lat91}).  
There are continuous efforts to provide nuclear data for supernovae 
with the advance of experimental and theoretical nuclear physics 
since the 1990s.  
Experimental data of unstable nuclei became available 
at radioactive nuclear beam facilities 
along with the advance of nuclear many body theories 
(\cite[Serot \& Walecka 1986]{Serot1986}, 
\cite[Brockmann \& Machleidt 1990]{Brockmann1990}).  
Systematic measurements of the radii of neutron rich isotopes 
are valuable inputs to constrain the interaction 
in the nuclear models for supernova EOS 
(\cite[Shen et al. 1998a, 1998b, 2011]{shen98a}\cite[]{shen98b}\cite[]{shen11}).  

The two EOS tables by Lattime-Swesty and Shen et al.\ have been popular 
in many  supernova simulation studies.  
There is recent progress of the EOS tables at low and high densities.  
One direction is extension to include exotic degrees of freedom 
such as hyperons and quarks at high densities 
(\cite[Ishizuka et al. 2008]{Ishizuka2008}, 
\cite[Sagert et al. 2009]{sagert09}, 
\cite[Nakazato et al. 2010a]{nak10a}).  
The other direction is extension to consider 
the mixture of nuclei at low densities 
(\cite[Hempel et al. 2012]{hempel12}).  
Recent observations of neutron star masses and radii are 
valuable to constrain the behavior of EOS at high densities.  
The observation of the massive neutron star of $\sim$2M$_{\odot}$ 
(\cite[Demorest et al. 2010]{Demorest2010}) 
is strong constraint to select the EOS sets for supernova simulations.  
Determination of the neutron star radii is also influential 
(\cite[Steiner et al. 2010]{Steiner2010}) 
and needs further careful analysis with model dependence 
(\cite[Suleimanov et al. 2011]{Suleimanov2011}).  

\subsection{Neutrinos in supernova dynamics}

An important key for the successful explosion is the counteracting 1\% effects.  
Neutrinos are playing a critical role to determine the energy budget 
in the supernova dynamics.  
Since the Fe core of $\sim$10$^{3}$ km collapses to the neutron star of $\sim$10 km, 
there is gravitational energy release, which amounts to $\sim$10$^{53}$ erg.  
The explosion energy including kinetic, internal and potential energies 
is $\sim$10$^{51}$ erg from observations.  
Although the explosion energy sounds just like 
a fraction of the gravitational energy release, 
the most of the energy are actually carried away as supernova neutrinos.  
The total energy of the emitted neutrinos amounts to $\sim$10$^{53}$ erg 
from the observation of the supernova neutrinos.  
Therefore, only 1\% is used for the explosion 
out of the total release of the emitted neutrinos.  
To understand this delicate role of neutrinos, 
careful evaluation of the neutrino-matter interaction 
is essential in every stage of the supernova dynamics.  

It is helpful to describe further some of the important stages 
as examples of the 1\% effects or the issue of 10$^{51}$ erg.  
At the early stage after the bounce, 
the initial shock wave stalls on the way 
because the energy is used up during the propagation.  
After the core bounce, 
the shock wave must propagate through the Fe core 
up to its surface against the falling material of outer layers.  
The energy loss occurs due to Fe dissociation during the propagation.  
Although the shock wave is launched up to $\sim$150 km, 
it stalls without success in the spherical simulations.  
Looking at the energy budget, 
the initial energy of the shock wave amounts to several 10$^{51}$ erg 
by the gravitational energy of inner core.  
The energy loss, on the other hand, due to the Fe dissociation 
amounts to minus several 10$^{51}$ erg.  
Since this loss uses up the initial energy, 
the hydrodynamical explosion in $\sim$100 ms is not possible.  

Additional assistance to bring the energy of 10$^{51}$ erg 
are necessary to revive the stalled shock wave.  
The neutrino heating mechanism behind the stalled shock wave 
is one of the plausible scenarios.  
At $\sim$100 ms after the core bounce, 
the stalled shock wave is hovering around 200 km 
with accretion of outer layers (Figure \ref{fig1}, left).  
At center, the proto-neutron star is born already 
and emits abundant neutrinos from its hot and lepton rich core.  
A part of them hit the material just below the shock wave and 
are absorbed by nucleons (and nuclei).  
This contributes to the heating through the energy transfer 
from the neutrinos to the matter.  
This heating energy amounts to several 10$^{51}$ ergs, 
which is comparable to the other factors described above.  
The size of this effect depends in detail on 
the shock wave dynamics, the energy and flux of neutrinos, 
the amount of the target material and duration time.  
In order to examine this process precisely, one has to 
solve the neutrino transfer together with the hydrodynamics.  
Modern numerical simulations must clarify 
whether the neutrino heating mechanism can revive 
the stalled shock wave as seen originally 
in the explosion by \cite{Bethe85} via the delayed explosion mechanism.  

\subsection{Multi-dimensions with supercomputing technology}

\begin{figure}[t]
\begin{center}
 \includegraphics[width=5cm]{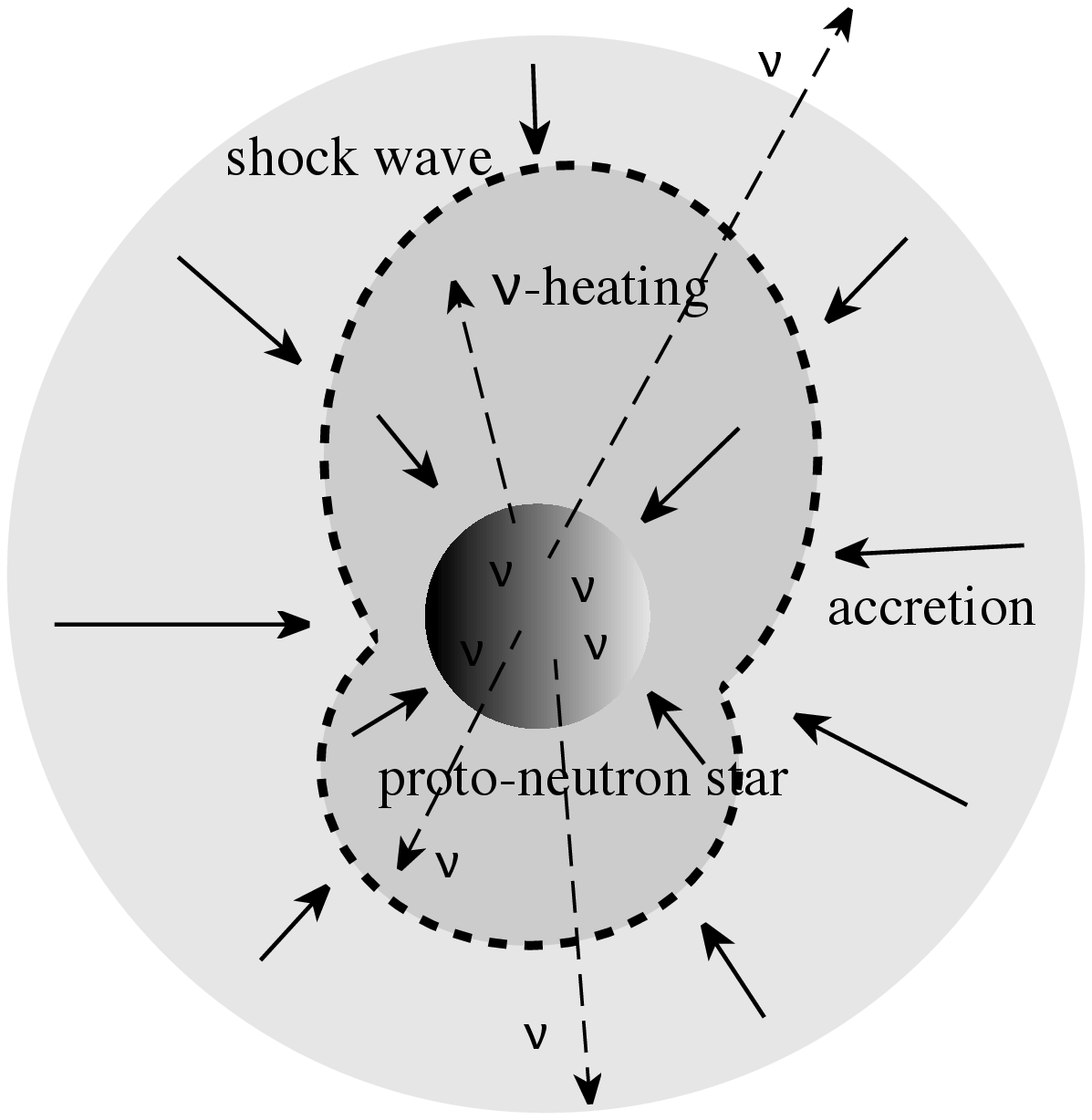} \hspace{0.2cm}
 \includegraphics[width=5.5cm]{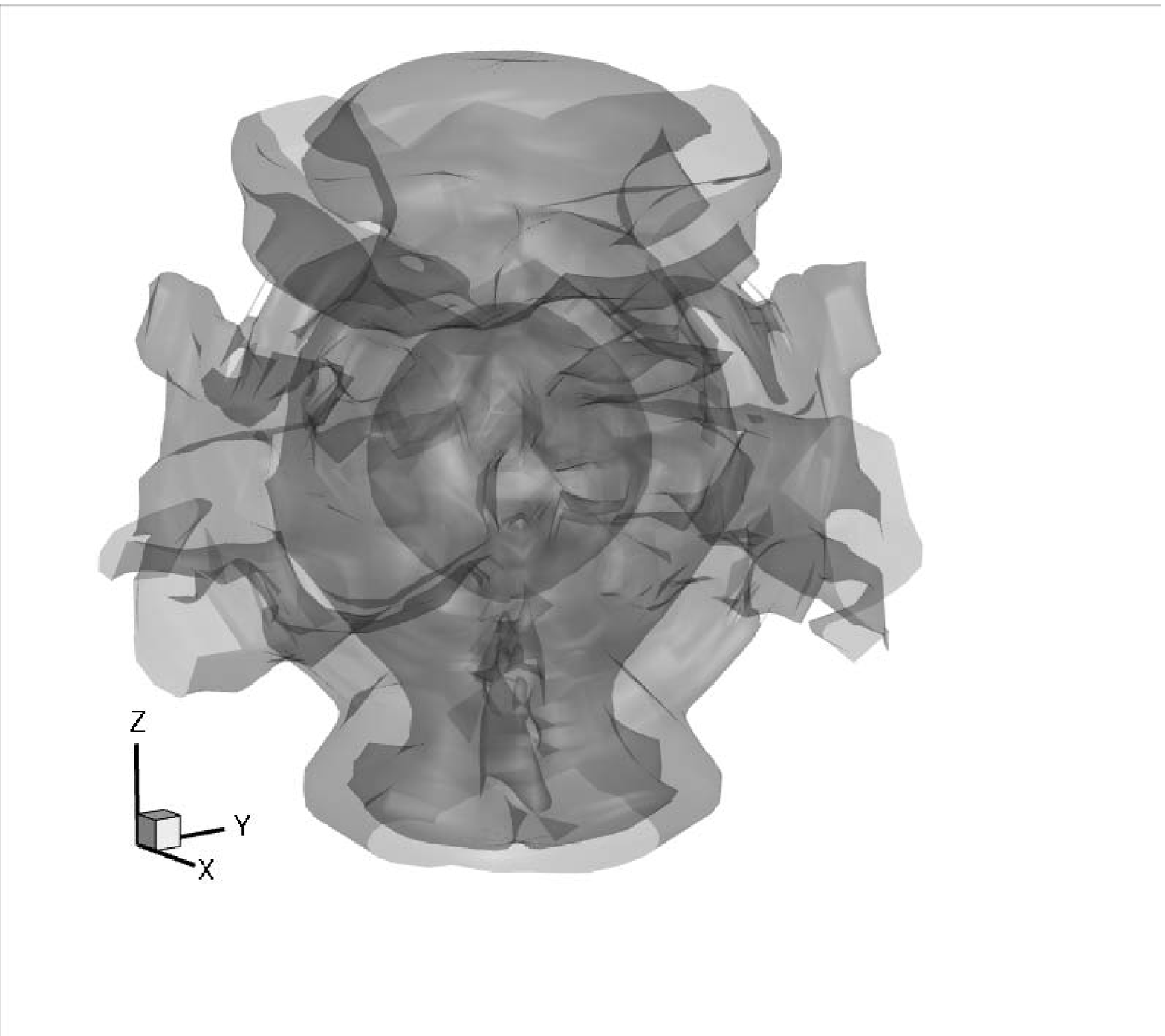} 
 \caption{(Left) Schematic diagram of the deformed shock wave 
 and neutrino heating.  
 (Right) Neutrino distributions in a supernova core evaluated 
 by neutrino transfer in 3D.  }
   \label{fig1}
\end{center}
\end{figure}

In addition to the role of neutrinos, novel features 
in multi-dimensional simulations are pivotal in the supernova mechanism.  
These are also challenging problems, which need the latest 
technology of supercomputers.  
When one breaks the symmetry from 1D (spherical) to 2D (axial) and 
3D in the numerical simulations, 
there are always new findings on the explosion mechanism 
due to hydrodynamical instabilities such as convections and 
standing accretion shock instabilities (SASI) 
(\cite[Blondin et al. 2003]{Blondin2003}).  
Moreover, the importance of asymmetry is supported by the observational facts 
on the asymmetric properties of supernova remnants.  
The main idea is the combination of the neutrino heating and 
the hydrodynamical instabilities.  
Because of the asymmetric launch of the shock wave 
(ex.\ up/down asymmetric), 
the shock wave along a certain direction reaches 
up to a larger distance than that realized in 1D.  
The neutrino heating below the deformed positions of 
the stalled shock wave helps to push the further propagation 
(Figure \ref{fig1}, left).  
The main gain is the sufficient amount of time for neutrino heating,
since the advection time for the accretion of material 
becomes longer than the 1D case.  

It is to be noted that solving neutrino radiation-hydrodynamics 
is necessary to clarify this problem.  
Radiation-hydrodynamics in multi-dimensions is generally 
a difficult problem in astrophysics and engineering.  
Along with the advance of supercomputers, there is step-by-step 
progress in the numerical treatment of neutrino-radiation hydrodynamics.  
In 1D (spherical symmetry), it is possible to perform 
the first-principle-type calculations 
(\cite[Rampp et al. 2000]{Rampp00}, 
\cite[Liebend{\"o}rfer et al. 2001]{Liebendorfer01}, 
\cite[Sumiyoshi et al. 2005]{Sumiyoshi05}).  
However, no explosion is found in the numerical simulations 
of various research groups for typical progenitors.  
In 2D, state-of-the-art simulations with detailed neutrino and 
nuclear physics produce a dozen of explosion cases 
(See Table 1 in \cite[Kotake 2012b]{kotake2012b}).  
However, they have been done with approximate treatment of 
neutrino transfer.  
Although the neutrino heating mechanism is promising 
in most of the 2D simulations, there are also 
plausible scenarios under discussion.  
In order to determine the most dominant effect, 
further systematic studies in 2D are necessary.  

Whether this kind of multi-dimensional effects 
remains important in 3D is the current topics in supernova studies.  
The advantage of 3D extensions over 2D models have been discussed 
in the systematic analysis by a simplified model 
(\cite[Nordhaus et al. 2010]{Nordhaus2010}) 
but remains controversial with new results 
(\cite[Hanke et al. 2012]{Hanke2012}).  
There will be also new instabilities, which is hidden in 2D 
but appear in 3D.  
With the rapid progress of supercomputing technology, 
it became recently to perform the first 3D simulation 
starting from the collapse of a massive star.  
Takiwaki et al.\ have studied the 3D core-collapse supernovae 
by the ray-by-ray approach with the IDSA approximation 
(\cite[Takiwaki et al. 2012]{taki12}).  
Explosions are found in 3D modeling with 3D motion of 
convection and deformed propagation of shock wave. 
However, it is still premature to conclude the intrinsic 3D effects 
since they have to perform more simulations with higher resolutions.  
One of the largest 3D simulations is running at the K-computer at Kobe, 
Japan, which has a peak speed of 
10 PetaFLOPS (the world's 2nd fastest machine as of June 2012).  

It is also necessary to replace approximate treatment of the neutrino transfer 
to the accurate one.  
It has been a formidable task to solve the neutrino-radiation transfer 
in 3D space since it is a 6D phase-space problem.  
Recently there is remarkable achievement to solve the 6D Boltzmann equation 
for neutrino distributions 
(\cite[Sumiyoshi \& Yamada 2012]{sumi12}) 
in supernova cores 
(Figure \ref{fig1}, right; see also \cite[Kotake et al. 2012a]{kotake2012a}).  
Future ExaFLOPS supercomputers will enable full 3D simulations 
of neutrino-radiation hydrodynamics of core-collapse supernovae.  

\section{Toward the birth of pulsars}

Until we get the solid mechanism for the successful explosion, 
one needs to assume explosions for astrophysical applications 
such as nucleosynthesis by parameterizing the shock wave  launch.
There are a number of applications to explore the neutron star 
properties after the supernova explosion.  

By putting initial condition of a proto-neutron star 
born in supernovae, 
it is possible to follow the thermal evolution of the proto-neutron star 
through the cooling by emission of neutrinos (e.g. \cite[Pons et al. 1999]{Pons99}).  
By predicting the neutrino spectra, average energies and luminosities 
as a function of time, one can explore the EOS influence as well as 
the dependence on the properties of proto-neutron stars.  
These are also important templates of supernova neutrino bursts 
for future detection of the next galactic supernova.  
The neutrino signals from the collapse of massive stars 
have been systematically studied 
including the case of black hole formations 
(e.g. \cite[Sumiyoshi et al. 2007]{Sumiyoshi2007}).  
Their evaluation of spectra has been applied to predict the event 
numbers (\cite[Nakazato et al. 2010b]{nak10b}) 
at the current and future facilities of neutrino detection 
(\cite[Kistler et al. 2011]{Kistler2011}).  

There have been studies of pulsar kicks and rotations 
through simulation of the parameterized explosion 
after the bounce (e.g. \cite[Blondin \& Mezzacappa 2007]{Blondin}).  
For example, 
\cite{Wongwathanarat2010} have studied the pulsar kicks and rotations 
by following the motion of accretion and ejecta, to predict 
the velocity of neutron stars through momentum conservation and 
gravitational pulls.  
They obtain the time evolution of pulsar kicks and angular momentum,
and find a moderate number of $\sim$300 km/s for kick velocities  
and $\sim$600 ms for spin periods.  
In principle, it is necessary to connect this kind of outcome 
with the progenitor mass and rotations via full supernova simulations.  

In summary, the study of core-collapse supernovae is fascinating 
with multi-physics from femto- to kilometer scales.  
It is a challenging problem regarding extreme conditions, 
counter-acting 1\% effects of neutrinos and multi-dimensions.  
Understanding the mechanism of supernova explosions requires 
neutrino-radiation hydrodynamics in three dimensions.  
Implementation of microphysics on nuclear data into the numerical simulation 
is mandatory.  
With the progress of 
supercomputing resources, there are promising 2-D cases of explosions.
The current challenge is the 3D numerical simulations 
to solve the neutrino-radiation hydrodynamics 
on the latest and future technology of supercomputers.  
These numerical challenges are indispensable for clarifying the essential 
cause of supernova dynamics, and to answer the birth of pulsars 
from the gravitational collapse of massive stars.  

\section*{Acknowledgements}
This article is based on the fruitful collaboration with 
H. Nagakura, 
K. Kotake, T. Takiwaki, 
K. Nakazato, H. Suzuki, 
and S. Yamada.  
The author acknowledges the usage of the supercomputers at 
Research Center for Nuclear Physics (RCNP) in Osaka University, 
The University of Tokyo, 
Yukawa Institute for Theoretical Physics (YITP) in Kyoto University 
and High Energy Accelerator Research Organization (KEK).  
The numerical study on core-collapse supernovae using the supercomputer 
facilities is supported by the HPCI Strategic Program of 
the Ministry of Education, Culture, Sports, Science and Technology 
(MEXT), Japan.  
This work is partially supported by the Grant-in-Aid for Scientific Research on Innovative Areas (Nos. 20105004, 20105005) 
and the Grant-in-Aid for the Scientific Research (Nos. 22540296, 24244036) 
from MEXT, Japan.  


\end{document}